\documentclass[12pt,preprint]{aastex}
\begin{document}
\title{ULTRA-COMPACT EMBEDDED CLUSTERS IN THE GALACTIC PLANE} 

\author{Michael J. Alexander\altaffilmark{1} and Henry A. Kobulnicky\altaffilmark{1}}

\altaffiltext{1}{Department of Physics and Astronomy, University of Wyoming, Laramie, WY 82071, USA}

\begin{abstract}  
We have identified a previously unrecognized population of very compact, embedded low-mass Galactic stellar clusters. These tight (r$~\approx~$0.14 pc) groupings appear as bright singular objects at the few arcsec resolution of the {\it Spitzer Space Telescope} at 8 and 24 $\mu$m but become resolved  in the sub-arcsecond UKIDSS images. They average six stars per cluster surrounded by diffuse infrared emission and coincide with 100 -- 300 M$_{\sun}$ clumps of molecular material within a larger molecular cloud. The magnitudes of the brightest stars are consistent with mid- to early-B stars anchoring $\sim$80 M$_{\sun}$ star clusters. Their evolutionary descendants are likely to be Herbig Ae/Be pre-main sequence clusters. These ultra-compact embedded clusters (UCECs) may fill part of the low-mass void in the embedded cluster mass function. We provide an initial catalog of 18 UCECs drawn from infrared Galactic Plane surveys.   
\end{abstract}


\section{Introduction}
Most, if not all, stars are born in stellar clusters. It has been estimated that 96\% of massive OB ($>$8 M$_{\sun}$) stars are associated with clusters \citep{dW05}. For nearby ($\lesssim$2 kpc) embedded star clusters, \citet{la03} found a flat mass distribution function, implying a power law ($\alpha=-2$) distribution by number for young clusters. Their relation exhibits a sharp turnover for clusters with total stellar masses less than $\sim$50 M$_{\sun}$ and suggests that $>$90\%\ of all stars form in clusters more massive than this lower limit. \citet{gu09} surveyed 36 star clusters in young star-forming regions (SFRs), mostly within 1 kpc of the sun, and found an average of 26 members per cluster with mean radii of 0.39 pc. These clusters are quite young, as evidenced by their high incidence of young stellar objects (YSOs). Despite recent advances and observations, the census of the smallest embedded clusters is still incomplete owing to the limited depth and angular resolution of large scale infrared (IR) surveys.

During a study of SFRs in the Galactic Plane \citep{al12}, we serendipitously identified two compact stellar clusters. These objects have pointlike or marginally resolved morphologies in the few arcsecond resolution mid-IR {\it Spitzer Space Telescope} images at [8.0] and [24] \micron, but are resolved in the sub-arcsecond $JHK$ images from the United Kingdom Infrared Deep Sky Survey (UKIDSS; \citealt[][]{lu08}). The mid-IR images are typically dominated by a single bright object that exhibits a steeply rising spectral energy distribution (SED) through the mid-IR. The putative clusters are often found within infrared dark clouds (IRDCs) surrounded by nebulosities tracing the hot dust, reflection nebulosity, and PAH emission characteristic of young embedded clusters. 

Using brightness and color criteria derived from the prototype sources, we searched for similar objects in the Galactic Legacy Infrared MidPlane Survey Extraordinaire (GLIMPSE; \citealt[][]{be03}) Point Source Catalogs (PSCs) as well as a subsample of massive YSO (MYSO) candidates from the {\it MidCourse Space Experiment} ({\it MSX}) Red MSX Source (RMS) catalog  \citep{ur11}. The search yielded additional candidates, and we present an initial (incomplete) sample of 18 ultra-compact embedded clusters (UCECs) and their properties.

\section{Identification of Candidate Clusters}
We used the mid-IR magnitudes of the prototype clusters to define two criteria to search for additional candidate clusters in the GLIMPSE PSCs. We required that sources have a red IR color\footnote{GLIMPSE PSCs include $JHK_{S}$ photometry from the Two Micron All Sky Survey (2MASS, \citealt[][]{sk06}).} ($K_{S} - $[3.6] $>$ 2) and a bright ($<$5 mag.) GLIMPSE  detection. Although the second criterion allows for any IRAC band, it is nearly always satisfied by [8.0] as the other bands saturate at fainter magnitudes. The search was limited to the GLIMPSE I area that overlaps with UKIDSS ($\approx$96 sq. deg. between $\ell=$15\degr\ -- 66\degr) to ensure the availability of high-resolution imaging. These criteria pick out 391  candidate clusters for which we examined UKIDSS images and tentively identified three types of objects: evolved stars, single YSOs, and potential clusters. We eliminated objects that appeared single and isolated, lacking apparent extended structures; these constituted the majority of the initial list and are most likely field giants or AGB stars. We retained six objects that exhibited multiple red stars with rising SEDs; these types of objects were invariably accompanied by an extended near-IR nebulosity, most prominently at $K$ band.

Literature searches on color-selected clusters revealed that several were identified previously as candidate MYSOs. Therefore, we expanded our search to include MYSO candidates from Table~3 of \citet{ur11}, which have estimated distances. The UKIDSS images revealed that some of the MYSO candidates appear to be small stellar clusters. The classification is ambiguous in some cases because of the high levels of diffuse emission associated with the objects. We found 12 objects that are candidate compact stellar clusters.   
 
Table~\ref{table} lists the 18 most compelling UCEC candidates. Column 1 is the ID number and columns 2 and 3 are Galactic longitude and latitude. Column 4 is the angular radius, r$_{c}$, estimated by-eye to include sources within the IR nebulosity. Column 5 is number of stars per cluster, N$_{*}$, with a membership probability greater than 75\%\ after background subtraction (discussed below). Column 6 is the LSR velocity of the peak $^{13}$CO emission from the Galactic Ring Survey \citep{ja06}, and column 7 is the near kinematic distance derived from the velocity and a rotation curve \citep{cl85}. Column 8 is the gas mass derived from the distance in column 7 and the 1.1 mm flux from the Bolocam Galactic Plane Survey (BGPS) Catalog using Equation 2 from \citet{ro10}, unless otherwise specified. Column 9 gives the physical radius, R$_{p}$, column 10 is the associated MSX ID \citep{ur11}, if applicable, and column 11 specifies if there is evidence for an associated compact H{\scriptsize II} region from the Multi-Array Galactic Plane Imaging Survey (MAGPIS, \citealt[][]{he06}) 6 or 20 cm maps. 

\section{Ultra-compact Embedded Clusters}
\subsection{General Properties}
\begin{figure}
\includegraphics[width=18cm]{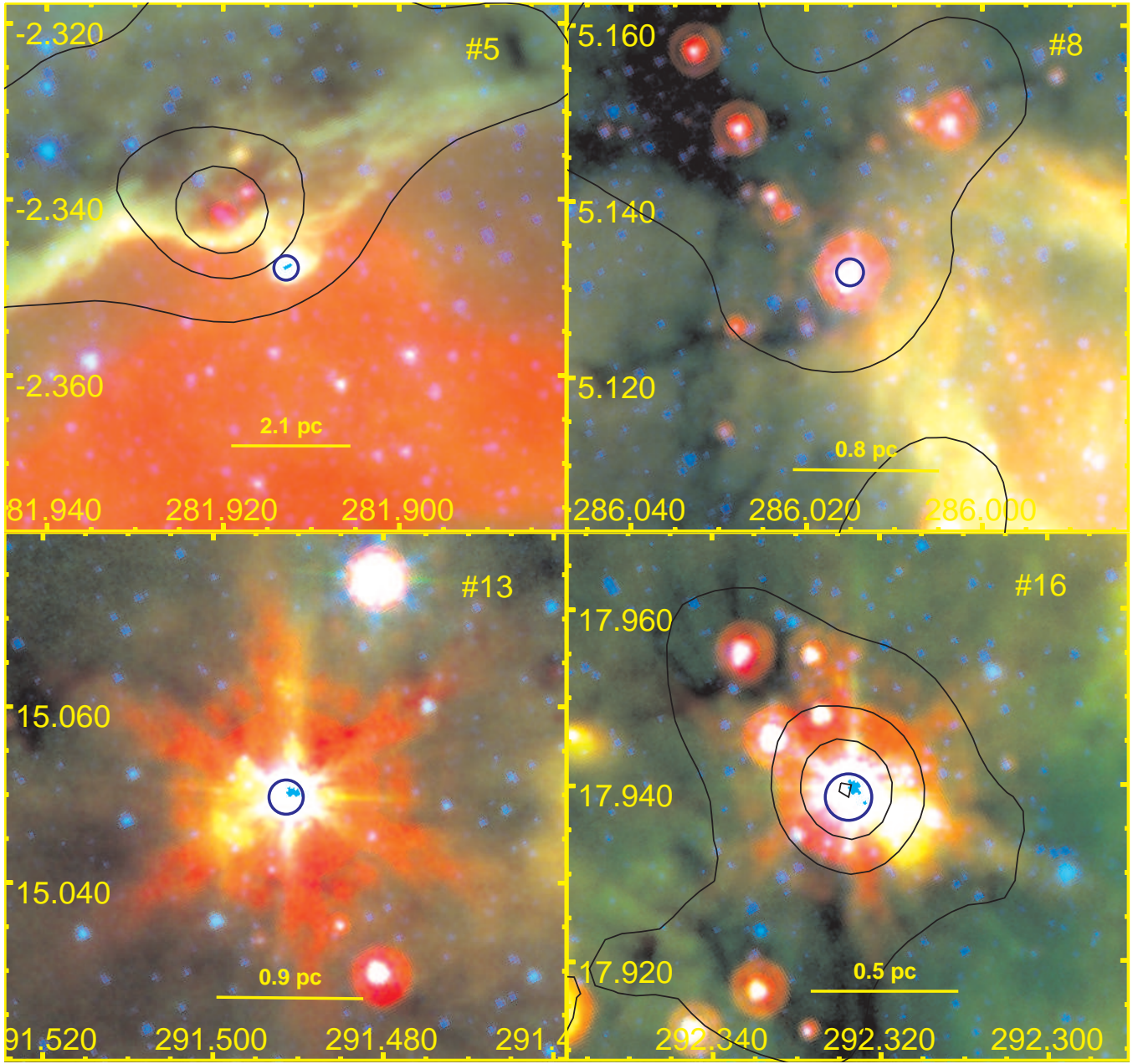}
\caption{Three-color [4.5], [8.0] and [24] (blue, green, red) image of four UCECs from Table~\ref{table}. They are \#5 (upper left), \#8 (upper right), \#13 (lower left), and \#16 (lower right). Blue circles indicate UCEC radii from Table~\ref{table} and the 1\arcmin\ yellow bar gives the linear scale at the adopted distance. Black contours outline BGPS 1.1 mm emission at 0.1, 0.6, 1.1, 1.6, and 2.1 Jy~beam$^{-1}$. The coordinates are Equatorial J2000. \label{glimpse}}
\end{figure}

Figure~\ref{glimpse} is a three-color ([4.5], [8.0], and [24]) image of four UCECs. Blue circles mark the locations of the UCECs and have radii given in Table~\ref{table}. The 1\arcmin\  yellow bar illustrates the linear size scale in pc at the adopted distance, and black contours outline BGPS 1.1 mm emission. UCEC \#13 (lower-left) falls outside the BGPS coverage so no contours are shown, however, IRDCs are apparent within the region. Most of the UCECs in Table~\ref{table} have associated IRDCs indicating that these regions are active, or future, SFRs and likely lie at the near kinematic distance. The IRDCs connected with the clusters contain a median gas mass of 141 M$_{\sun}$; the millimeter continuum maps indicate that they are generally part of larger molecular clouds. 

The prototype UCECs are $\sim$5 -- 11$\arcsec$ in radius and appear pointlike at [8.0] and [24], while the [3.6] and [4.5] images may partially resolve the clusters into 2 -- 4 highly blended sources. The average IRAC magnitudes for the brightest point source in an individual UCEC are 7.9, 7.3, 5.7, and 4.9 for [3.6], [4.5], [5.8], and [8.0], quite close to the saturation limits. The magnitudes demonstrate a rising SED typical of YSOs and indicate the extreme youth and/or embedded nature of these objects. Although the cluster radius is large compared to the 1\farcs2 pixel size of IRAC, the combination of source brightness and high source density makes it difficult or impossible to resolve individual stars, and it is not obvious that the sources contain compact clusters. 

We estimated field star contamination by comparing $H$ vs. $H - K$ cluster color-magnitude diagrams (CMDs) with field CMDs using UKIDSS DR7 photometry\footnote{http://surveys.roe.ac.uk/wsa/}. The cluster radius, $r_c$, defined the target area, and an annulus from $r_c$ to 33\arcsec\ was the field area. The field should be large enough to provide good statistics yet small enough to represent the local field population and extinction. The cluster and field CMDs were divided into bins and compared, yielding the membserhip probability, P = (N$_{C}$ - A$\times$N$_{F}$)/N$_{C}$, where N$_{C}$ is the number of target stars a given bin, N$_{F}$ is the field star count, and A is target-to-field area ratio. We used 729 bin size and center combinations to mitigate biases caused by any particular binning strategy. This is similar to the procedure used by \citet{ma10}, except that high extinction towards UCECs reduces J-band source counts so we are limited to H- and K-band. This technique  estimates where a cluster lies in color-magnitude space but cannot tell if an individual star is a member because stars are selected statistically. A true assignment of cluster membership generally requires stellar spectra to assign a spectral type and distance.

\begin{figure}
\includegraphics[width=18cm]{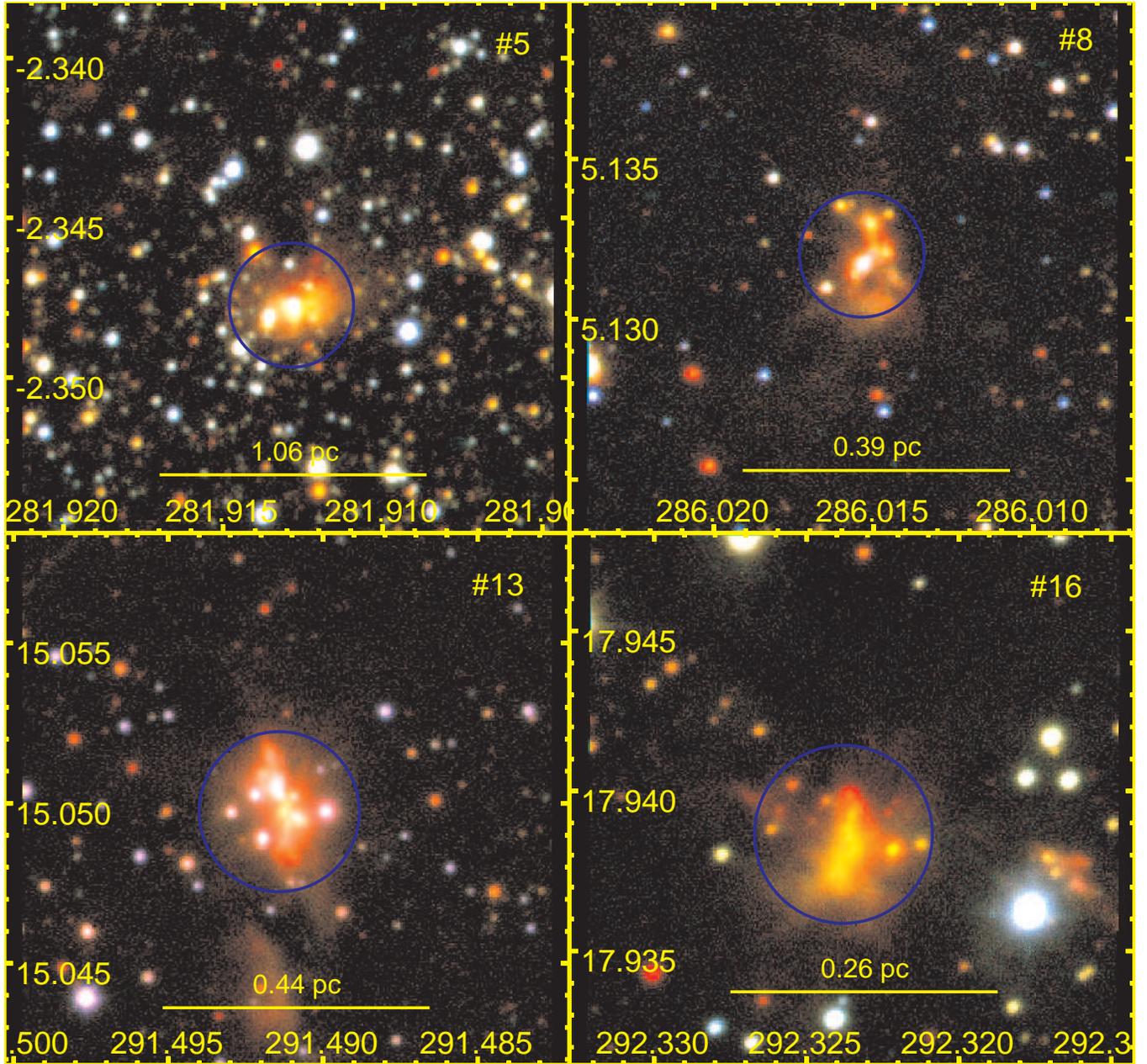}
\caption{Three-color $JHK$ (blue, green, red) image of the UCECs from Figure~\ref{glimpse}. Blue circles incidate the cluster aperatures. The 30\arcsec\ yellow bar shows linear scale. \label{ukidss}}
\end{figure}

Figure~\ref{ukidss} shows zoomed three-color UKIDSS $JHK$ (blue, green, and red) images of the four UCECs, as in Figure~\ref{glimpse}, where the yellow bar is now 30\arcsec. The cluster nature of these sources is more pronounced at sub-arcsecond resolution, but there is still some source blending. The UCECs exhibit structured diffuse emission, likely arising from hot dust and/or reflection nebulosity. These sources span distances from 2 -- 7 kpc, but have similar physical sizes. 

\subsection{UCEC \#5}
UCEC \#5 (Figure~\ref{ukidss}, upper-left) shows the highest field star density, consistent with its large distance of 7.3 kpc. Figure~\ref{cmd} (upper-left) is an $H$ vs $(H - K)$ CMD for \#5 showing cluster members ($>$75\%\ probability) as pluses and field stars as grey dots. The solid and dotted lines are a zero-age main sequence (ZAMS) isochrone \citep{ma08} and a 0.5 Myr pre-main sequence isochrone (PMS, \citealt[][]{si00}), respectively. Both have been placed at the estimated cluster distance (7.3 kpc) and extincted according to the reddening vector \citep{ca89}. The diamond, circle, and asterisk on the MS isochrone mark the location of an A0, B2, and O9 star, respectively. The molecular cloud associated with \#5 has a radial velocity V$_{LSR}~=~$103.9 km~s$^{-1}$ and lies near the tangent point in this direction, making the distance unambiguous. The extinction value was estimated by-eye to match the locus of putative cluster members. The spread in $H - K$ is likely caused by strong differential reddening within the cluster, so cluster members are not expected to lie along any single isochrone. The two brightest sources may be O stars, if they are single, while the rest fall in the early-B range. 

UCEC \#5 protrudes from the edge of a bright-rimmed cloud (Figure~\ref{glimpse}, upper left) likely formed by a known H{\scriptsize II} region \citep{lo89}. The rim of PAH and molecular material is traced by the 1.1~mm continuum contours and may indicate the presence of a swept-up shell of material making the cluster a candidate for triggered star formation (SF). The millimeter emission suggests a total molecular gas mass of 5400 M$_{\sun}$ for the entire clump. However, the emission does not peak over the cluster, so we estimate an upper limit of 130 M$_{\sun}$ for the gas mass within a 9\arcsec\ radius of the UCEC. The cluster itself does not exhibit a peak in either the 6 or 20~cm maps from MAGPIS (2\arcsec\ and 6\arcsec\ beam FWHM, respectively), but diffuse radio continuum from the main H{\scriptsize II} region procludes an accurate upper limit.

\subsection{UCEC \#8}
UCEC \#8 was the first UCEC identified and appears in the upper-right panel of the figures. Figure~\ref{glimpse} shows the UCEC sandwiched between an IR-bright bubble (N74, \citealt[][]{ch06}) and the edge of an IRDC. \citet{ja06} place the IRDC at the near distance of 2.7 kpc, which we adopt for the UCEC. The upper-right panel of Figure~\ref{cmd} shows the cleaned CMD, with symbols as before. The eight sources within this region display a similar color and suggests that they occupy the same volume. Most of the sources are again consistent with early-B stars, taking into account differential reddening. The 1.2 mm maps presented by \citet{ra06} have higher resolution (11\arcsec\ FWHM) than the BGPS (33\arcsec\ FWHM), so we adopt their gas mass estimates of 1792 M$_{\sun}$ for the entire IRDC and 73 M$_{\sun}$ for the peak coincident with the UCEC.

\subsection{UCEC \#13\label{u13}}
Figure~\ref{glimpse} (lower-left) shows the extreme mid-IR brightness of UCEC \#13, as well as nearby IRDCs. Figure~\ref{ukidss} reveals the cluster nature of the object and shows many bright sources embedded in diffuse emission. The CMD for this cluster (Figure~\ref{cmd}, lower-left panel) shows six sources consistent with B star magnitudes at approximately $A_{V}~=~$15 mag. UCEC \#13 falls outside the BGPS coverage but has a detection from the James Clerk Maxwell Telescope (JCMT) at 450 \micron\ \citep{di08}. We use Equation 9 from \citet{ro10} to calculate the total gas mass of the cloud, adopting 3.0 kpc for the distance, 6.7 cm$^{2}~$g$^{-1}$ for the dust grain opacity \citep{os94}, and a dust-to-gas ratio of 0.01. This yields a total mass for the region of 217 M$_{\sun}$.

\subsection{UCEC \#16}
UCEC \#16 (Figure~\ref{glimpse}, lower-right) is situated at the center of an IRDC complex. The bright core is saturated in the mid-IR images, and several bright point sources, likely YSOs, are scattered throughout the IRDC. Figure~\ref{ukidss} shows that all sources within the cluster radius are extremely red. This is evident in the CMD (Figure~\ref{cmd}) which shows all five of the stars have $(H - K)~>~2$ and implies $A_{V}~>~30$. At the estimated distance and reddening, they are consistent with B stars. This region lies at a distance of 1.8 kpc and has a total gas mass of 124 M$_{\sun}$ \citep{ra06}.

\section{Discussion}
Our initial search shows that young, embedded compact clusters can be selected by an IR color and magnitude cut with follow-up visual inspection. Undoubtedly, there are additional compact clusters that remain unidentified because of very high extinction, large field star densities, extreme compactness, or incompleteness stemming from high diffuse background levels. Our analysis revealed that all 12 of the UCECs from the MYSO sample were missing either a 2MASS $K_{S}$ or [8.0] catalog entry. The missing detection at $K_{S}$ is probably from source blending, while at [8.0] saturation likely kept the object out of the point source catalogs. Therefore, this color selection technique is limited to sources faint enough to be unsaturated in GLIMPSE. If we were to relax the selection criteria to include fainter sources with $[8.0]~=~$5--6 mag, the number of selected objects increases to more than a thousand. These could include even lower-mass clusters and those too faint to be included in the MSX catalog, but visual classification of such a large sample is beyond the scope of this work.

Sub-arcsecond mid-IR imaging was performed on 14 MYSO candidates at 24.5 \micron\ \citep{dw09} and on 346 MYSOs at 10.3 \micron\ \citep{mo07}. These studies found that approximately 20 -- 25\%\ of MYSO candidates have multiple detections and/or extended diffuse emission. These wavelengths primarily detect stars with strong IR excesses rather than stellar photospheres and may miss sources that lack or have only a weak excess. However, their results are consistent with a portion of MSYO candidates being compact stellar clusters.

\begin{figure}
\includegraphics[width=18cm]{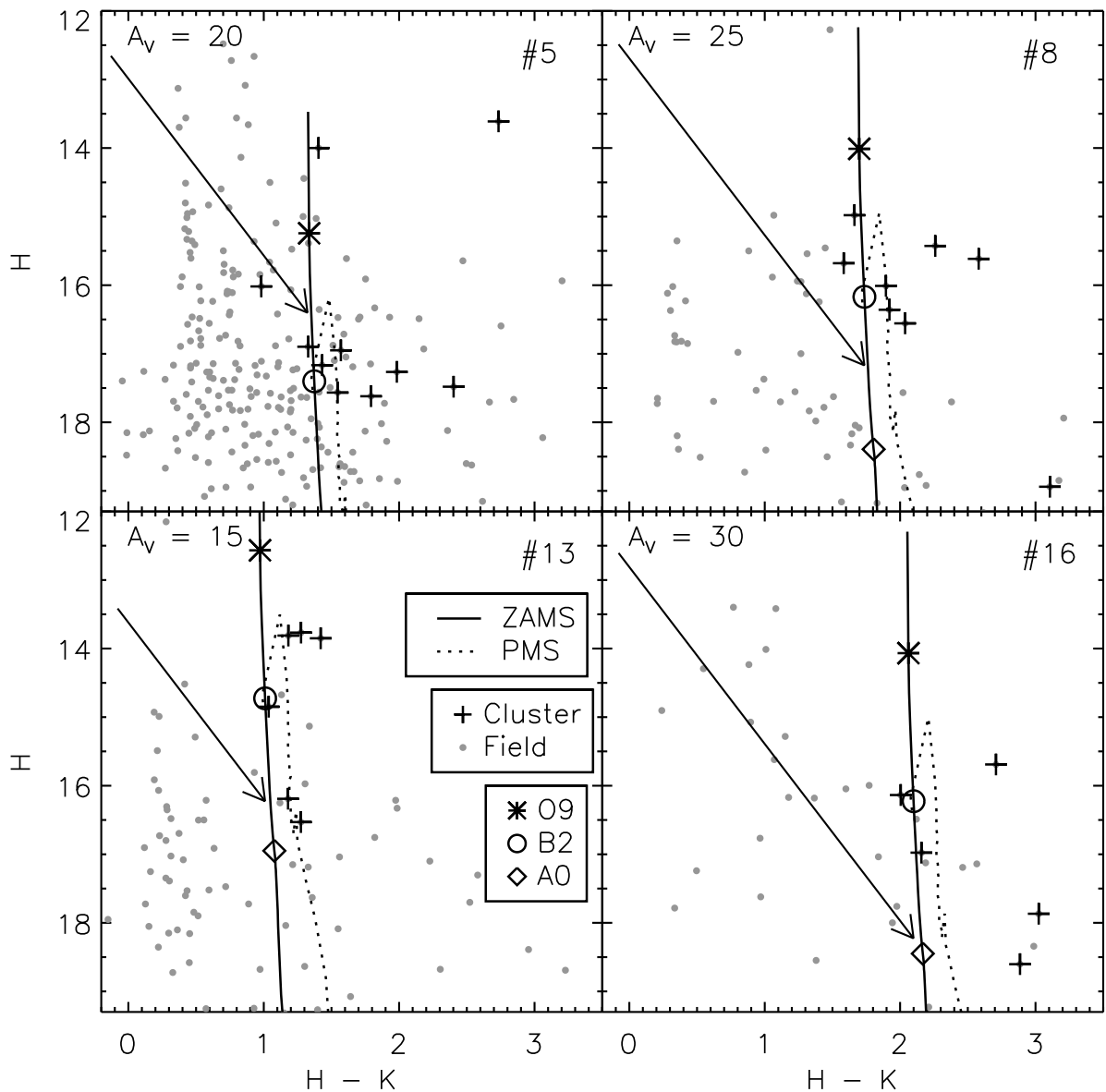}
\caption{CMDs for the UCECs from Figure~\ref{glimpse}. Dots represent stars within the field annulus, between $r_{C}$ and 33\arcsec, and pluses mark sources within the cluster aperture that survive the field star subtraction ($>$75\% probability). The solid and dotted lines are ZAMS and 0.5 Myr PMS isochrones at the cluster distance and extincted according to the reddening vector. \label{cmd}}
\end{figure}

Only four of 18 UCECs have possible radio detections. UCEC \#11 was detected at 3.6 and 1.3 cm by \citet{sa08} and they estimate a spectral type of B2 -- B3 for the exciting source. Three others (\#4, \#5, and \#6) appear to be associated with faint 20 cm emission from the MAGPIS survey \citep{he06}, however they do not appear in the MAGPIS point source catalog \citep{wh05}. The estimated completeness limit at 20~cm is 14~mJy, which is sensitive enough to detect an O9.5V at 10~kpc based on the Lyman continuum flux \citep{ma05}. This indicates that the most massive star(s) within the majority the UCECs is an early-B star producing relatively few Lyman continuum photons, consistent with the inferences drawn from the CMDs in Figure~\ref{cmd}. Another possibility is that the stars earlier than B0 are so young that they have not yet formed a detectable H{\scriptsize II} region \citep{ur11}. In some cases a single, bright source appears to dominate the UCECs and may in fact be a MYSO, but in others the IR flux is more evenly distributed among cluster members, in which case the most massive star may be an early- to mid-B star. 

\citet{te97} identified small clusters of PMS stars around Herbig Ae/Be stars. These clusters have radii of about 0.2 pc, typically contain 4 -- 12 stars, and $<$few Myr old. In these clusters the maximum stellar mass is correlated with the K band source counts. Field-star-subtracted source counts (6--8 stars) from Figure~\ref{cmd} are roughly consistent with those found in Herbig Be PMS clusters (2--16) \citep{te99}. It is likely that UCECs suffer a higher level of extinction owing to their highly embedded nature and, as a result, source counts within UCECs may be underestimated compared to more exposed and evolved Herbig Ae/Be clusters. This evidence suggests that UCECs represent a younger, more heavily embedded phase  destined to evolve into Herbig Ae/Be clusters after a few Myr.

\citet{we10} found a correlation between the most-massive cluster member (m$_{max}$) and the total cluster mass (M$_{ecl}$) that cannot be explained by a random sampling of the IMF. The m$_{max}$-M$_{ecl}$ relation is incomplete below 100 M$_{\sun}$ \citep{we10}, but is supported by later investigations \citep{ki11}. An accurate determination of M$_{ecl}$ depends on cluster age because of the increasing probability of losing cluster members over time \citep{bo03}. UCECs are ideal objects for further probing the m$_{max}$-M$_{ecl}$ relation because they are likely to have M$_{ecl}~<~100$ M$_{\sun}$ and young enough to have not lost a significant number of cluster stars.

Figure~\ref{cmd} indicates that m$_{max}$ is near an $\sim$8 M$_{\sun}$ B2V star, which implies M$_{ecl}~\sim~80$ M$_{\sun}$, while an O9V star ($\sim$20 M$_{\sun}$) and a B8V ($\sim$4 M$_{\sun}$) would have M$_{ecl}$ of 251 and 26 M$_{\sun}$, respectively \citep{we10}. If UCECs are 80 M$_{\sun}$ clusters the SF efficiency, SFE = M$_{*}$/(M$_{*}$~+~M$_{gas}$) would be 0.52, 0.27, and 0.39 for \#8, \#13, and \#16, respectively, while the median gas mass of the entire sample (141 M$_{\sun}$) produces a SFE of 0.36. These SFE values are slightly higher, though still consistent, with studies of other Galactic clusters and SFRs \citep{la03,al07}. The CMDs in Figure~\ref{cmd} show that the clusters may contain more than one early-B star, which would increase the implied number of unseen low-mass stars for a standard IMF. These sources are absent either because they fall below the UKIDSS detection limit or the IMF is truncated in these types of objects.

After stars form within a cluster, they immediately begin to expel the surrounding ISM. The rapid explusion of gas alters the cluster's potential well and may cause clusters to dissolve \citep{bo03}. \citet{la03} estimate up to 95\%\ of embedded clusters will disperse in under 5 -- 10 Myr, and those that do survive longer typically have masses over 500 M$_{\sun}$. This puts an upper limit on the lifetime of UCECs ($<$ a few Myr) and suggests that they will quickly disperse. Such clusters, in any case,  would be difficult to identify after a few Myr once the large IR luminosity arising in circumstellar and intracluster dust diminishes.

\citet{ja10} suggest that IRDCs are the precursors to massive stars and star clusters. The presence of UCECs embedded within IRDCs supports this hypothesis. After several Myr, it is likely that the IRDC will have dissipated and SF ceased. Small clusters, including UCECs, will be distrupted and may appear as a single loose cluster or stellar association of a few hundred solar masses. Thus, large stellar associations may be comprised of the distributed remnants of many smaller clusters born out of the same IRDC. The gas-free merger of small clusters may explain why the m$_{max}$-M$_{ecl}$ relation differs from random IMF sampling (for clusters $>$100 M$_{\sun}$) by limiting accretion and growth of the most massive members, except in the most massive molecular clouds \citep{we10}.

UCECs may represent an unrecognized but significant population of low-mass stellar clusters destined to quickly disperse into the Galactic stellar field. In large enough numbers, these types of objects may be numerous enough to steepen the low-mass end of the embedded cluster mass function.

\acknowledgments We wish to thank our anonymous referee for many helpful comments. We also thank Charles Kerton and Ed Churchwell for their suggestions. MJA is supported by NASA GSRP fellowship NNX 10-AM10H. HAK is supported by NASA grant ADAP-NNX10AD55G.

\clearpage
\begin{deluxetable}{rcccccccccc}
\tablecaption{Candidate UCECs \label{table}}
\tablewidth{14.5 cm}
\tabletypesize{\scriptsize}
\tablehead{
\colhead{ID}& \colhead{l}& \colhead{b}& \colhead{r$_{c}$}& \colhead{N$_{*}$}& \colhead{V$_{LSR}$}& \colhead{D}& \colhead{M$_{gas}$}& \colhead{R$_{p}$}& \colhead{RMS}& \colhead{Radio} \\
\colhead{ }& \colhead{\degr}& \colhead{\degr}& \colhead{\arcsec}& \colhead{}& \colhead{km~s$^{-1}$}& \colhead{kpc}& \colhead{M$_{\sun}$}& \colhead{pc}& \colhead{ID}& \colhead{Continuum} \\
\colhead{(1)}& \colhead{(2)}& \colhead{(3)}& \colhead{(4)}& \colhead{(5)}& \colhead{(6)}& \colhead{(7)}&  \colhead{(8)}& \colhead{(9)}& \colhead{(10)}& \colhead{(11)}
}
\startdata
 1&18.3706&-0.3825& 8&10& 43.9&3.7& 187&0.14&G018.3706-00.3818& no \\
 2&21.2372& 0.1940& 9& 6& 25.7&2.2&  65&0.10&\nodata          & no \\
 3&24.6346&-0.3236& 6& 4& 42.7&3.2& 141&0.10&G024.6343-00.3233& no \\
 4&28.8621& 0.0653& 6& 3&102.8&6.8&3770&0.20&G028.8621+00.0657&maybe \\
 5&30.4108&-0.2283& 7&10&103.9&7.3&$<$130&0.25&G030.4117-00.2277&maybe \\
 6&30.8191& 0.2730& 7& 3& 97.7&6.6& 164&0.22&G030.8185+00.2729&maybe \\
 7&34.7120&-0.5952& 6& 3& 44.4&2.9& 166&0.08&G034.7123-00.5946& no \\
 8&38.9369&-0.4588& 7& 8& 40.7&2.7&  73\tablenotemark{a}&0.09&G038.9365-00.4592& no \\
 9&39.4946&-0.9939& 9& 7& 53.6&3.8&\nodata&0.17&G039.4943-00.9933& no \\
10&40.3062&-0.4313& 6& 7& 74.3&5.8& 176&0.17&\nodata          & no \\
11&42.0981& 0.3515& 5& 9& 23.0&1.6&  10&0.04&\nodata          &yes \\
12&43.9954&-0.0121& 7& 4& 65.4&5.3& 333&0.18&G043.9956-00.0111& no \\
13&50.2212&-0.6068& 9& 6& 39.3&3.0& 217\tablenotemark{b}&0.13&G050.2213-00.6063& no \\
14&52.9193&-0.8608& 6& 2& 56.0&5.1&\nodata&0.15&\nodata          & no\\
15&52.9216&-0.4889& 7& 5& 45.0&4.2&  85&0.14&G052.9221-00.4892& no \\
16&53.1410& 0.0697&10& 5& 21.6&1.8& 124\tablenotemark{a}&0.09&G053.1417+00.0705& no \\
17&53.2194& 0.0485&10& 4& 23.9&1.9&  17\tablenotemark{a}&0.09&\nodata          & no \\
18&59.3601&-0.2061&11& 6&\nodata&2.3\tablenotemark{c}& 208\tablenotemark{c}&0.15&\nodata  & no \\
\enddata
\tablenotetext{a}{\citet{ra06}}
\tablenotetext{b}{Calculated in Section~\ref{u13}.}
\tablenotetext{c}{\citet{bi10}}
\end{deluxetable}

\clearpage

\clearpage

\clearpage


\begin{thebibliography}{}
\bibitem[Aguirre et al.(2011)]{ag11} Aguirre, J.~E., Ginsburg, A.~G., Dunham, M.~K., et al.\ 2011, \apjs, 192, 4 
\bibitem[Alexander et al.(2012)]{al12} Alexander, M.~J., et al.\ 2012, in prep.
\bibitem[Alves et al.(2007)]{al07} Alves, J., Lombardi, M., \& Lada, C.~J.\ 2007, \aap, 462, L17 
\bibitem[Benjamin et al.(2003)]{be03} Benjamin, R.~A., Churchwell, E., Babler, B.~L., et al.\ 2003, \pasp, 115, 953 
\bibitem[Billot et al.(2010)]{bi10} Billot, N., Noriega-Crespo, A., Carey, S., et al.\ 2010, \apj, 712, 797 
\bibitem[Boily \& Kroupa(2003)]{bo03} Boily, C.~M., \& Kroupa, P.\ 2003, \mnras, 338, 665 
\bibitem[Cardelli et al.(1989)]{ca89} Cardelli, J.~A., Clayton, G.~C., \& Mathis, J.~S.\ 1989, \apj, 345, 245 
\bibitem[Churchwell et al.(2006)]{ch06} Churchwell, E., et al.\ 2006, \apj, 649, 759 
\bibitem[Clemens(1985)]{cl85} Clemens, D.~P.\ 1985, \apj, 295, 422 
\bibitem[de Wit et al.(2005)]{dW05} de Wit, W.~J., Testi, L., Palla, F., \& Zinnecker, H.\ 2005, \aap, 437, 247 
\bibitem[de Wit et al.(2009)]{dw09} de Wit, W.~J., Hoare, M.~G., Fujiyoshi, T., et al.\ 2009, \aap, 494, 157 
\bibitem[Di Francesco et al.(2008)]{di08} Di Francesco, J., Johnstone, D., Kirk, H., MacKenzie, T., \& Ledwosinska, E.\ 2008, \apjs, 175, 277 
\bibitem[Gutermuth et al.(2009)]{gu09} Gutermuth, R.~A., Megeath, S.~T., Myers, P.~C., et al.\ 2009, \apjs, 184, 18 
\bibitem[Helfand et al.(2006)]{he06} Helfand, D.~J., Becker, R.~H., White, R.~L., Fallon, A., \& Tuttle, S.\ 2006, \aj, 131, 2525 
\bibitem[Jackson et al.(2006)]{ja06} Jackson, J.~M., et al.\ 2006, \apjs, 163, 145 
\bibitem[Jackson et al.(2010)]{ja10} Jackson, J.~M., Finn, S.~C., Chambers, E.~T., Rathborne, J.~M., \& Simon, R.\ 2010, \apjl, 719, L185 \bibitem[Kirk \& Myers(2011)]{ki11} Kirk, H., \& Myers, P.~C.\ 2011, \apj, 727, 64 
\bibitem[Kroupa(2001)]{kr01} Kroupa, P.\ 2001, \mnras, 322, 231 
\bibitem[Lada \& Lada(2003)]{la03} Lada, C.~J., \& Lada, E.~A.\ 2003, \araa, 41, 57 
\bibitem[Lockman(1989)]{lo89} Lockman, F.~J.\ 1989, \apjs, 71, 469 
\bibitem[Lucas et al.(2008)]{lu08} Lucas, P.~W., Hoare, M.~G., Longmore, A., et al.\ 2008, \mnras, 391, 136 
\bibitem[Maia et al.(2010)]{ma10} Maia, F.~F.~S., Corradi, W.~J.~B., \& Santos, J.~F.~C., Jr.\ 2010, \mnras, 407, 1875 
\bibitem[Marigo et al.(2008)]{ma08} Marigo, P., Girardi, L., Bressan, A., et al.\ 2008, \aap, 482, 883 
\bibitem[Martins et al.(2005)]{ma05} Martins, F., Schaerer, D., \& Hillier, D.~J.\ 2005, \aap, 436, 1049 
\bibitem[Mottram et al.(2007)]{mo07} Mottram, J.~C., Hoare, M.~G., Lumsden, S.~L., et al.\ 2007, \aap, 476, 1019 
\bibitem[Ossenkopf \& Henning(1994)]{os94} Ossenkopf, V., \& Henning, T.\ 1994, \aap, 291, 943 
\bibitem[Rathborne et al.(2006)]{ra06} Rathborne, J.~M., Jackson, J.~M., \& Simon, R.\ 2006, \apj, 641, 389 
\bibitem[Rosolowsky et al.(2010)]{ro10} Rosolowsky, E., Dunham, M.~K., Ginsburg, A., et al.\ 2010, \apjs, 188, 123 
\bibitem[S{\'a}nchez-Monge et al.(2008)]{sa08} S{\'a}nchez-Monge, {\'A}., Palau, A., Estalella, R., Beltr{\'a}n, M.~T., \& Girart, J.~M.\ 2008, \aap, 485, 497 
\bibitem[Siess et al.(2000)]{si00} Siess, L., Dufour, E., \& Forestini, M.\ 2000, \aap, 358, 593 
\bibitem[Skrutskie et al.(2006)]{sk06} Skrutskie, M.~F., Cutri, R.~M., Stiening, R., et al.\ 2006, \aj, 131, 1163 
\bibitem[Testi et al.(1997)]{te97} Testi, L., Palla, F., Prusti, T., Natta, A., \& Maltagliati, S.\ 1997, \aap, 320, 159 
\bibitem[Testi et al.(1999)]{te99} Testi, L., Palla, F., \& Natta, A.\ 1999, \aap, 342, 515 
\bibitem[Urquhart et al.(2011)]{ur11} Urquhart, J.~S., Moore, T.~J.~T., Hoare, M.~G., et al.\ 2011, \mnras, 410, 1237 
\bibitem[White et al.(2005)]{wh05} White, R.~L., Becker, R.~H., \& Helfand, D.~J.\ 2005, \aj, 130, 586 
\bibitem[Weidner et al.(2010)]{we10} Weidner, C., Kroupa, P., \& Bonnell, I.~A.~D.\ 2010, \mnras, 401, 275 
\end{thebibliography}
\end{document}